\documentstyle{mn2e}
\input psfig

 \def\be{\begin{equation}}
 \def\ee{\end{equation}}
\def\solmas{{M$_\odot$}}
\def\simless{\mathbin{\lower 3pt\hbox
   {$\rlap{\raise 5pt\hbox{$\char'074$}}\mathchar"7218$}}}   % < or of order
\def\simgreat{\mathbin{\lower 3pt\hbox
   {$\rlap{\raise 5pt\hbox{$\char'076$}}\mathchar"7218$}}}   % > or of order
\def\etal{{\rm et al.}}

\def\solmas{{M$_\odot$}}
\def\solm{{M_\odot}}

\def\solr{{R_\odot}}
\def\au{ AU}
\def\Mb{M_{\rm bin}}
\def\Rb{R_{\rm bin}}
\def\Rp{R_{\rm peri}}
\def\Rs{R_{\rm semi}}

\def\mnras{{MNRAS}}

\def\ARAA{{ARA\&A}}

  \makeatletter
  \ifx\CUP@mtlplain@loaded\undefined  % this is a font switch for CUP
  \makeatother  
    
  \else
    
  \fi

\loadboldmathitalic % loads bold math italic
\loadboldgreek      % loads bold greek
  
\makeatletter
\ifx\CUP@mtlplain@loaded\undefined  % this is a font switch for CUP
  \newfont\bit{cmbxti10 at 9pt}
\else
  \newfont\bit{mtbxti10 at 9pt}
\fi
\makeatother

\def\LaTeX{L\kern-.36em\raise.3ex\hbox{a}\kern-.15em
    T\kern-.1667em\lower.7ex\hbox{E}\kern-.125emX}

\newcommand{\gsim}{\mathrel{\hbox{\rlap{\lower.55ex \hbox {$\sim$}}
                   \kern-.3em \raise.4ex \hbox{$>$}}}}
\newcommand{\lsim}{\mathrel{\hbox{\rlap{\lower.55ex \hbox {$\sim$}}
                   \kern-.3em \raise.4ex \hbox{$<$}}}}

\title[Massive binary stars] {Binary systems and stellar mergers in massive star formation}
\author[I. A. Bonnell \etal]
  {Ian A. Bonnell$^1$\thanks{E-mail: iab1@st-and.ac.uk} and  Matthew R. Bate$^2$ \\
$^1$ School of Physics and
  Astronomy, University of St Andrews, North Haugh, St Andrews, Fife,
  KY16 9SS. \\
$^2$ School of Physics, University of Exeter, Stocker Road, Exeter, EX4 4QL \\ }

\date{\today}

\begin{document}

\maketitle

\begin{abstract}

We present a  model for the formation of high-mass close binary systems in the context of forming massive stars through gas accretion in the centres
of stellar clusters. A low-mass wide binary evolves under mass accretion towards
a high-mass close binary, attaining system masses of order 30-50 \solmas at separations
of order 1 \au. The resulting high frequency of binary systems with two massive components is
in agreement with observations. These systems are typically highly eccentric and may evolve to have periastron separations less than their stellar radii. Mergers of these binary systems are therefore likely and can lead
to the formation of the most massive stars, circumventing the problem of radiation pressure  stopping the accretion. The stellar density required to induce binary mergers is $\approx 10^{6}$ stars pc$^{-3}$, or 
$\approx 0.01$ that required for
direct stellar collisions.

\end{abstract}


\section{Introduction}

The formation of high-mass stars is a large unknown in modern astronomy.
While our understanding of the formation of low-mass stars has improved dramatically over the
past decade (Larson~2003), we still do not know whether massive star formation is basically a scaled-up version of low-mass star formation or if it results from a dramatically different process (Stahler, Palla \& Ho 2000; Bally \& Zinnecker 2005). The basic questions concern
how mass is added and how it interacts with the large radiation pressure from the high stellar luminosity
once the star reaches masses in excess of $10 \solm$. Models of scaled-up low-mass star formation
have repeatedly shown that if the gas contains typical interstellar dust, then the radiation pressure
deposits sufficient momentum into the dust  (well coupled to the gas) to repel the infalling gas
(Yorke \& Kr\"ugel 1977; Wolfire \& Casinelli 1987; Beech \& Mitalas 1994; Edgar \& Clarke 2004).

Potential solutions to this problem include accretion through a disc (Yorke \& Sonnhalter 2002), accretion overpowering the
radiation pressure (McKee \& Tan 2003), and  massive star formation through mergers of (dust free) stars (Bonnell, Bate \& Zinnecker~1998; Bonnell \& Bate 2002). Disc accretion can help 
in two ways. Firstly, the matter accretes through a small solid angle and is thus less exposed
to the radiation pressure. Secondly, as young stars are generally thought to be rapidly rotating,
the star can be significantly cooler at the equator than at its poles. Thus, there is less radiation
pressure in the equatorial regions to oppose the accretion. Using this last factor in their models
as well as  a sophisticated treatment of the radiation transfer, Yorke \& Sonnhalter (2002) showed that
it is possible to accrete up to masses of the order 30 \solmas\ before the radiation pressure repels
the infalling envelope. McKee \& Tan (2003) have speculated that a very concentrated centrally condensed core will provide large accretion rates onto a forming massive star which will then
overpower the radiation pressure of the growing star. There are two potential limitations
to this process. Firstly, such a core, even if centrally condensed and supported by turbulence, is
very likely to fragment and form a small stellar cluster (Dobbs, Bonnell \& Clark 2005). Secondly, the
implied accretion rates are similar to the models of Yorke \& Sonnhalter (2002) which show that
the infalling envelope is indeed repelled once the star attains masses of order 30 \solmas.

The third potential solution involves the merger of lower-mass stars in a dense stellar cluster
(Bonnell \etal 1998). While certainly the most exotic of the three, it does have the attraction that
any dust will be destroyed in the lower-mass stars and thus there is no problem with the radiation
pressure from the forming massive star. The obvious difficulties with this model is that it
requires very high stellar densities (of order $10^8$ stars pc$^{-3}$) in order for stellar collisions
to be sufficiently common to form massive stars in less than $10^6$ years. Although such densities
are $10^3$ times higher than generally found in the cores of young stellar clusters, they are not inconceivable as the dynamics of accretion can induce the core of a cluster to
contract significantly (Bonnell \etal 1998; Bonnell \& Bate 2002).  

One inescapable feature of massive stars is that they form in rich stellar clusters (Clarke \etal~2000; Lada \& Lada 2003).  Even apparently isolated massive stars
are best explained as being runaways from stellar clusters (de Wit \etal 2005; Clarke \& Pringle 1992). These systems have
a regular field-star IMF and thus we need to put massive star formation into the context of forming
many more low-mass stars.  Models for the formation of stellar clusters, neglecting any feedback processes, show that
the fragmentation of a turbulent molecular cloud can form hundreds of stars and that they
follow a field-star like IMF (Bonnell, Bate \& Vine 2003). Intriguingly, the initial masses of these
stars, even those that end up being high-mass, are all initially close to the Jeans mass of the cloud (Bonnell, Vine \& Bate  2004). The
development of high-mass stars in these systems occurs due to subsequent accretion 
onto the forming cluster, where the stars near the centre of the potential accrete more rapidly
and thus attain much higher masses (Bonnell \etal 2001a; Bonnell \etal 2004). This effect is increased once a significant disparity in masses is achieved. It is this competitive accretion which explains
the origin of the IMF while maintaining a low median stellar mass in the cluster (Bonnell \etal 2001b;, Bonnell \etal 2003). Massive star formation in the context of low-mass star formation can thus be
explained as being primarily due to the environment of the forming massive star (Bonnell \etal 2004),
linking the formation of a massive star to the formation of a stellar cluster. However we still need
to understand how the mass accretion overcomes the radiation pressure to form the most massive stars.

A further complication to any model of massive star formation is that most massive stars are
in binary systems (Mason \etal 1998; Preibish \etal 1999; Garcia \& Mermilliod 2001). 
These systems are often very close, with
separations $\simless 1$ AU, and companions that are also high-mass stars (Garmany, Conti \* Massey 1980; Bonanos \& Stanek 2005). It is these 
close massive binary systems that motivate the present study.  In Section~2 we explain the difficulties
of forming close binary systems. In Section 3 we discuss the numerical simulations of cluster formation.
Section~4 explains the formation of massive binary systems in stellar clusters and  their  orbital evolution.  The potential for binary mergers is discussed in Section~5. We discuss the implications of these results in \S~6 and finally, our conclusions are given in Section~7.

\section{The problem: forming close  binary stars}

Forming close binary stars systems is difficult even amongst lower-mass stars, If the binary components form through fragmentation (eg, Boss 1986; Bonnell 1999), then the Jeans radius at the
point of fragmentation must be smaller than the binary separation,
\be
\Rb \simgreat  2 R_J \propto T^{1/2} \rho^{-1/2}.
\ee
This implies a
high gas density and thus a low Jeans mass,
\begin{equation}
M_* \approx M_J \propto T^{3/2} \rho^{-1/2}.
\ee
This results in the mass of the individual stars being directly related to their separation,
\be
\Rb \propto \frac{M_*}{T},
\end{equation}
such that close systems have very low masses (Boss~1986).  For example, if the typical 30 AU binary has solar mass
components, then a $1/3$ AU binary should have components of $0.01$ \solmas  (Bonnell \& Bate 1994).
Forming close binary stars in situ is therefore difficult as
it requires subsequent accretion to reach stellar masses (Bate~2000).
An alternative is that the components form at greater separation and then
are brought together.
Recent simulations of low-mass star formation in a cluster environment have shown that close binaries can result
from the induced evolution of wider systems (Bate, Bonnell \& Bromm 2003a). The binaries
evolve due to gas accretion, angular momentum lost to circumbinary discs (e.g.  Pringle ~1991;Artymowicz \etal 1991) and dynamical interactions with other stars. Can the same
processes explain high-mass close binary systems?

\subsection{Accretion and Binary Evolution}   %%% Second level section head (remove "%" symbol)

Accretion onto binary systems has the potential of forming close systems out of
wider systems at the same time as forming higher-mass components. In order to see this, let
us consider the angular momentum of a binary system,
\be
L \propto \Mb^{3/2} \Rb^{1/2}.
\ee
If the accreted material has zero net angular momentum, as is expected if it infalls
spherically symmetric, then $L\approx$ constant, and the binary separation should be a strong function of the mass,
\be 
\Rb \propto \Mb^{-3}.
\ee
If instead, the accreted material has constant specific angular momentum, the same as the
initial binary, then the total angular momentum will scale with the mass of the binary, $L \propto M$ and
thus the orbital separation will scale with the mass as
\be
\Rb \propto \Mb^{-1}.
\ee
In a turbulent medium, as expected here,  the angular momentum
of each parcel of  infalling gas is essentially randomly oriented. The net angular momentum will 
then do a random walk with increasing mass such that
\be
L \propto \Mb^{1/2},
\ee
and thus
\be
\Rb \propto \Mb^{-2}.
\ee
Under these basic assumptions, we can see that accretion onto a binary system can significantly
decrease its separation at the same time as it increases its mass (Bate \& Bonnell~1997; Bate~2000).

\section{Calculations}
The numerical simulations discussed here were first reported in Bonnell, Bate \& Vine~(2003).
% with an additional higher resolution rerunning of part of the original simulation to check
% on the effects of the numerical resolution. 
The simulations  used the Smoothed Particle Hydrodynamics (SPH) method (Monaghan~1992). The
code has variable smoothing lengths in time and in space and solves for the
self-gravity of the gas and stars  using a tree-code (Benz \etal~1991).  The initial conditions
consisted of 1000 \solmas\ of gas in a $0.5$ pc radius uniform density sphere.
The cloud also contains significant turbulent motions such that the kinetic energy
is equal to the magnitude of the potential energy. The gas is isothermal at 10K which
implies an initial Jeans mass of $1 \solm$.
Star formation is modeled by the inclusion of sink-particles (Bate, Bonnell \& Price 1995)
that interact only through self-gravity and through gas accretion. Sink-particle creation occurs when 
dense clumps   of gas have  $\rho \simgreat 1.5 \times 10^{-15}$ g cm$^{-3}$, are
self-gravitating, and are contained   in a region such that the SPH smoothing lengths are smaller
than the  'sink radius' of  $200$ \au. This ensures that the initial sink-particles can have
masses as low as our resolution limit of $0.1 \solm$.
Gas particles are accreted if they fall within a sink-radius (200 \au) of a sink-particle and are bound to it.
In the case of overlapping sink-radii, the gas particle is accreted by the sink-particle to which
it is most bound.
The gravitational forces between sink-particles is smoothed within $160$ \au\ using the SPH kernel.
Thus any  binary separation within 160 \au\ is an overestimate of the true separation. 
This implies that the frequency of dynamical interactions in the simulation is artificially higher than would be the case
if the binary was allowed to evolve to smaller separations. This is offset to some degree by the
corresponding decrease in the strength of each interaction due to the gravitational softening.
The simulations were carried out  on the United Kingdom's Astrophysical Fluids Facility (UKAFF), a 128 CPU SGI Origin 3000 supercomputer.

\subsection{Reconstructing the binary's orbital parameters}

In spite of the fact the gravitational forces are smoothed at distances less than 160 \au, we
can still hope to extract information as to what the true binary separations should be. This
is possible as the orbital parameters depend solely on the system's mass, angular momentum and total energy, which are all directly calculable from the simulation.  Thus we can estimate what the true binary semi-major axis, $\Rs$, would be
in the absence of any gravitational smoothing as
\be 
\Rs = \frac{J^2}{G\Mb},
\ee
where  $J$ is the specific angular momentum, $M$ is the total mass of the binary system and $G$ is
the gravitational constant.  We can likewise calculate the eccentricity, $e$, of the orbit from
\be
e = \sqrt{1 + \frac{2 E J^2}{G^2\Mb^2}},
\ee
where $E$ is the total specific energy (energy per unit mass) of the binary system. This term
includes the smoothed gravitational potential that corresponds to the smoothed gravitational forces.
We can then estimate the semi-major axis of the orbit as well as the periastron separation,
\be
\Rp = \Rs (1-e), 
\ee
of the binary and determine its evolution. Furthermore, we can compare the periastron separation  to the stellar radii to estimate if the binary system would merge to form a more massive star.

\section{ Formation of high-mass binary stars}

The birth of high-mass stars in the simulation of Bonnell \etal (2003) is due to competitive accretion in a cluster environment. The simulation forms numerous sub-clusters  which eventually merge
to form one large system.  The high-mass stars form in the centres of the dense sub-clusters due to the combined 
potential that funnels the gas down to the centre of the system there to be accreted by the 
growing proto-massive star. The clusters themselves form as small-N groupings and grow by accreting stars
and infalling gas. Thus, the system is initially small and 3-body capture occurs readily (Binney \& Tremaine 1987). 
In small-N 
systems a central binary forms when three initially unbound stars 
pass so close that there is violent exchange of energy, with one 
star being ejected at high speed and the other two becoming bound 
to one another.  In larger systems this does not work as the larger velocity dispersion and smoother gravitational potential drastically reduces the probability of having a third star sufficiently close 
to extract the excess kinetic energy. 
As the high-mass stars all form in the centre of the 
individual  sub-clusters,  they generally have undergone three-body capture early in the cluster's growth and are therefore in binary systems.

\begin{figure}
%\vspace{-0.5truein}
\centerline{\psfig{figure=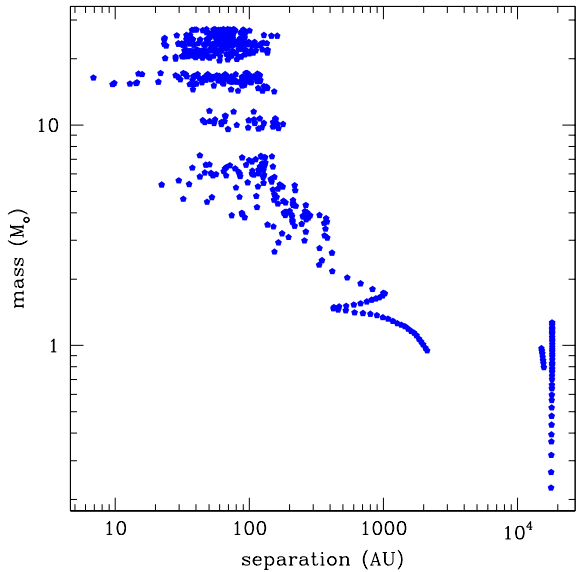,width=8.truecm,height=8.truecm}}
\caption{\label{binevol30}  The evolution of the binary separation is plotted against system mass for one of the high-mass binaries formed in the simulation. The system forms as a low-mass wide binary due to three-body capture and evolves to a high-mass closer binary due to gas accretion. The separation is an overestimate of the actual
separation due to the gravitational smoothing at separations less than 160 \au.}
\end{figure}

Once the binary has formed, continuing accretion onto it increases the masses of the individual stars.
Of equal importance is the effect of the accretion on the binary's separation. As the infalling gas has no correlation with the binary, its specific angular momentum is uncorrelated with that of the binary's. Thus, accretion does not significantly increase the binary's angular momentum and the separation of the
binary decreases as mass is added (see also Bate~2000). This process is illustrated in Figure~\ref{binevol30}
which shows the evolution of the binary mass versus orbital separation for one of the high-mass binary systems formed in the simulation (final masses 17 and 10 \solmas). The system originates as
a low-mass wide binary and evolves towards higher masses and smaller separations. Once the
separations are $\le 160$ \au, the gravitational smoothing stops the binary from evolving to
smaller separations. 

The early stages of the evolution are well parameterised by $\Rb\propto \Mb^{-2}$ 
as expected for accretion from a turbulent medium where each gas parcel has randomly
oriented angular momentum (see \S 2.1 above). Unfortunately, the binary's orbital evolution quickly enters the regime where the gravitational forces are smoothed (160 \au).  From this
point on, we
use the evolution of the energy and angular momentum to determine the evolution of the binary's orbit.

\subsection{ Evolution of the binary parameters}

\begin{figure}
%\vspace{-0.5truein}
\centerline{\psfig{figure=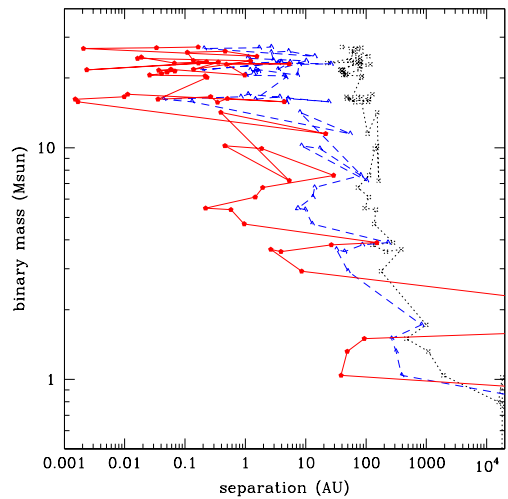,width=8.truecm,height=8.truecm}}
\caption{\label{binevol30semiperi} The binary's system mass is plotted against the deduced semi-major axis  (dashed line) and periastron
separation (solid line)  for the same high-mass binary as
in Figure~\ref{binevol30}. For comparison, the separation due to the gravitational smoothing is plotted as the dotted line. Note that the evolution is from low-mass wide system to a high-mass close system
with $\Rs\simless 10$\au\  and $\Rp\simless 1$ \au.}
\end{figure}

A binary's orbital parameters depend solely on the angular momentum, energy and mass of the binary system (See \S 3 above). Thus, as long as we can calculate these quantities, we can extract the
binary properties and their evolution. Figure~\ref{binevol30semiperi} shows the evolution
of the binary's system mass as a function of its semi-major axis and  periastron separation, for the
same binary as shown in Figure~\ref{binevol30}. The evolution of the smoothed separation is
also shown for comparison. The semi-major axis of the binary is generally much smaller than  the
smoothed separation. Thus, from separations of order 100 \au\ at masses of several solar masses, the
binary evolves to a semi-major axis of order 1 \au\ by the time the system has accreted up to 30 \solmas.
This once again roughly agrees with an $\Rb \propto \Mb^{-2}$ evolution as expected for
accretion from a turbulent medium. 
 
The evolution of the periastron separation is also plotted in Figure~\ref{binevol30semiperi}.
from this figure we can see that the binary is generally in a highly eccentric orbit where the 
periastron separation is an order of magnitude smaller than the semi-major axis, corresponding to an eccentricity of $e\approx 0.9$.  At one point the periastron separation appears to be larger
than the semi-major axis and this denotes a binary system which is unbound. At other points the
evolution decreases in system  mass indicating an exchange has occurred with a third star, or that
this third star temporarily passes closer to the primary. There is obviously some ambiguity in the evolution due to the gravitational smoothing imposed which maintains an artificially large
cross section for further dynamical interactions in addition to smoothing out these interactions. 
Nevertheless, we see that the periastron separation decreases  more dramatically than does the semi-major axis
and attains separations of order the size of high-mass stars ($\Rp \simless 0.05$ \au). 
In fact, at several points, the deduced
semi-major axis is smaller than the stellar radii of the stars. Such systems can reasonably be expected to 
undergo stellar mergers or in the very least, tidal forces which will reduce their semi-major axes
to be of order their periastron separations.

\section{Binary systems and stellar mergers}

\begin{figure}
%\vspace{-0.25truein}
\centerline{\psfig{figure=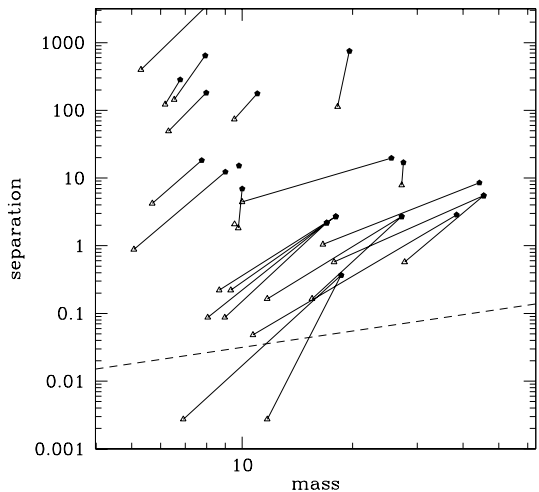,width=8.truecm,height=8.truecm}}
\caption{\label{binseps}  The binary  semi-major axis is plotted against binary mass (filled pentagons)
for all the systems containing at least one star with $m\ge 5 \solm$. The periastron separations
(open triangles) are also plotted for these systems against the individual component masses, and are joined to their total 
binary masseand semi-major axis by solid lines. Of note is that some binary systems have only
one individual component plotted indicating that the other star has another, closer companion, The dashed line indicates an approximation of the stellar radii  and thus periastron separations below this line would force the binary to merge.}
\end{figure}

The above evolution of one of the high-mass binaries is a typical example of the evolution
of the orbital parameters under mass accretion.  The distribution
of binary separation as a function of mass is shown for all the high-mass binary systems 
(where at least one star has $m\ge 5 \solm$) in Figure~\ref{binseps}. 
This figure plots the binary's semi-major axis
against the total binary mass (filled pentagons) and the individual component masses against the
periastron separations (open triangles), linked by solid lines for each system. The more massive stars
are generally in binaries with semi-major axes less than 10 \au\ and periastron separations less than 1 \au.
In fact, of the 10 stars at the end of the simulation with masses $m\ge 10\solm$,
all ten are in binary systems $\Rs \le 100$ \au\ and 8 of them have $\Rs \le 10$ \au. For the 18 stars between
5 and 10 \solmas, the numbers are 7 with $\Rs\le 100$\au\ and 5 with $\Rs\le10$\au.
 
In addition to the large fraction of close binary systems amongst the high-mass stars, it is also worth noting that generally both components of the binary are high-mass stars. Both of these properties
are in agreement with observations of massive binary systems (Mason \etal 1998;  Garcia \& Mermilliod~2001; Garmony \etal 1980). Of even greater potential significance is that the periastron separations of the systems are dangerously close to the estimated
size of the stellar radii (assuming a main-sequence mass-radius relationship), 
 \be
 R_* \approx \solr \left(\frac{M_*}{\solm}\right)^{0.8}.
 \ee
 Thus, many of these systems could be undergoing mergers,  
 while at least one system has  $\Rp < R_*$.
 In fact, this is only an instantaneous picture of the binary systems and as we can see from Figure~\ref{binevol30semiperi}, the evolution is somewhat chaotic and frequently perturbed by passing stars. 
 
\begin{figure}
%\vspace{-0.25truein}
\centerline{\psfig{figure=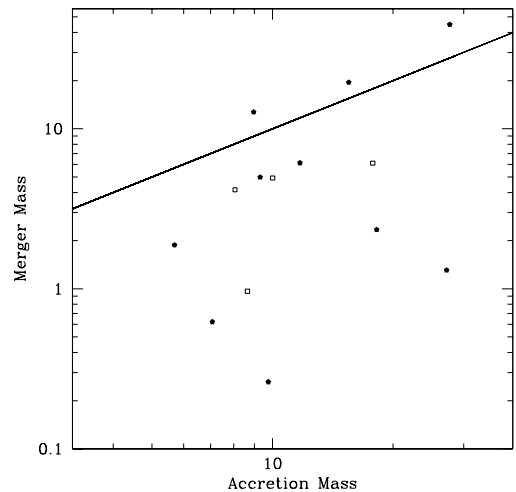,width=8.truecm,height=8.truecm}}
\caption{\label{accmerge} The mass gained in binary mergers, based on having periastron separations
smaller than one half of the primary's radius, is plotted against the mass acquired through gas accretion. The final products are plotted as filled pentagons while open squares denote merger products which subsequently merged with a more massive star. The mergers
are estimated by calculating the binary properties throughout the simulation and ensuring that each
star can only merge once with a more massive star. The solid line denotes equal mass being acquired through mergers and accretion. }
\end{figure}

 In order to quantify the frequency and relevance of binary mergers in high-mass star formation, we have 
 evaluated throughout the simulation when a binary system has a periastron passage closer than 
 half its stellar radii. Using this and the condition that each star can only merge with a more massive star once (only the most massive pre-merger star continues as an independent entity), we have estimated the potential for mass growth via binary mergers. This merger-acquired mass is plotted  in Figure~\ref{accmerge}
 against the mass acquired through gas accretion. The stars that survive the merger process are
 plotted as filled squares while those that merged during the evolution are plotted as open squares.
 We note from this figure that while most of the more massive stars should undergo mergers,
 these mergers are generally with lower-mass stars and thus would not significantly increase the stellar masses. This is not the case for at least
 three of the stars which would more than double their masses through mergers. In particular, the most
 massive star would increase its mass from $\approx 30\solm$ to nearly $75\solm$ through multiple binary mergers.  Thus, in
 some cases binary mergers could be a significant factor in forming massive stars.
  
\section{Discussion}

The potential for binary systems to merge and form the most massive stars is an intriguing
solution to the difficulty of overcoming the radiation pressure of the forming high-mass star.
A binary system with a semi-major axis of a few \au\ can be easily perturbed by an encounter with
a third star that passes near the binary. This encounter radius is significantly
larger than the size-scale for direct collisions and thus drastically reduces the requirement of
an ultradense cluster. The encounter time for an interaction at $R_{\rm enc} $ is given by
\begin{equation}
{1 \over t_{\rm enc}} = 16 \sqrt{\pi} n v_{\rm disp}R_{\rm enc}^2 \left(1+ {G\Mb \over 2 
 v_{\rm disp}^2 R_{\rm enc} }\right),
\end{equation} where $v_{\rm disp}$ is the velocity dispersion and $n$ is the  stellar density.
If we consider an eccentric binary of semi-major axis 1 \au, the
stellar density $n$ of order $10^6$ stars pc$^{-3}$ is required to have another star pass within $R_{\rm enc} \approx 2$\au, of order the apastron separation,  in $ t_{\rm enc} \le 10^6$ years, assuming  $v_{\rm disp}$ in the range of a few to 10 km s$^{-1}$. This is still a fairly high density
but not drastically so relative to  that of the cores of young stellar clusters (McCaughrean \& Stauffer 1994), especially considering 
that they could have been much denser in their earlier evolutions (Kroupa, Hurley \& Aarseth 2001; Scally, McCaughrean \& Clarke 2005). Central stellar densities an order of magnitude higher
are deduced in the more massive clusters near the Galactic centre (Figer \etal\ 1999).

An important implication of this work is that the most massive stars are more likely to be single stars,
or at least not in a close binary system.
This is due to their being merger products and thus the binary system is destroyed in forming this
very massive star. Intriguingly, well determined masses for stars in binary systems rarely exceed $\approx 50$ \solmas (Bagnuolo \etal 1992; Rauw \etal 1996; Massey \etal 2002; Bonanos \etal 2005)
with the most massive stars in a binary being a pair of $\approx 70$ \solmas (Rauw \etal 2004; Bonanos \etal 2004). This is in contrast with deduced
masses for single stars that can extend to $150$ \solmas and beyond (Davidson \& Humphreys 1997; Figer \etal 1998; Eikenberry \etal  2004).  This apparent discrepancy can be resolved either by
the spectroscopic masses being far in excess of the dynamical masses, or, by the most massive
stars having formed from the merger of two massive stars in a close binary system. 
 
\section{Conclusions}

The formation of high-mass close binary systems is a natural outcome of forming 
massive stars through gas accretion in a clustered environment.  Competitive
accretion in clusters naturally forms closer, more massive binaries out of lower-mass wide systems.
These systems can evolve from solar-mass binaries with separations of order 100 \au\ to
masses of $10-30 \solm$ with separations of order 1\au\ or less.  The frequency of high-mass
binary systems formed in a numerical simulation of a forming stellar cluster is 100 per cent
for stars with $m\ge 10\solm$ and  60 per cent for stars with $m\ge 5\solm$. The high
frequency of binary systems and the fact that they generally comprise two high-mass stars is in
agreement with observations (Mason \etal 1998; Garcia \& Mermilliod 2001; Preibish \etal 1999). 
The binaries 
typically have very eccentric orbits such that their periastron separations can be comparable to, or less than, the stellar radii. Mergers of these systems are therefore expected and can be a significant
factor in forming the most massive stars. Stellar densities required to perturb such a binary
and force it to merge are estimated to be of order $10^6$ stars pc$^{-3}$. 

\section*{Acknowledgments}
 We thank Hans Zinnecker for continually asking about binary
systems, the referee Ant Whitworth for comments which improved the text, and the organisers of the Massive Stars in Interacting Binaries meeting in Sacacomie, Que, which prompted this study. The computations reported here were performed using the U.K. Astrophysical Fluids Facility (UKAFF). MRB is grateful for the support of a Philip Leverhulme Prize.


\begin{thebibliography}{}

\bibitem[\protect\citeauthoryear{Artymowicz et 
al.}{1991}]{1991ApJ...370L..35A} Artymowicz P., Clarke C.~J., Lubow S.~H., 
Pringle J.~E., 1991, ApJ, 370, L35 
 
\bibitem[]{} Beech M., Mitalas R., 1994, ApJS, 95, 517

\bibitem[]{} Bagnuolo W. G., Gies D. R., Wiggs M. S., 1992, ApJ, 385, 708

\bibitem[]{} Bally J., Zinnecker H., 2005, AJ, in press

\bibitem[]{} Bate M. R., Bonnell I. A., Price N. M., 1995, \mnras, 277, 362.

\bibitem[]{} Bate M.~R., Bonnell I.~A., Bromm V., 2002, \mnras, 336, 705 

\bibitem[]{} Bate M.~R., Bonnell I.~A., 1997, \mnras, 285, 33

\bibitem[]{} Bate M.~R., 2000, \mnras, 314, 33

\bibitem[]{} Binney, J., \& Tremaine, S., 1987, Galactic Dynamics, Princeton U. Press 

\bibitem[\protect\citeauthoryear{Bonanos \& 
Stanek}{2005}]{2005ASPC..332..257B} Bonanos A.~Z., Stanek K.~Z., 2005, 
ASPC, 332, 257 
 
\bibitem[\protect\citeauthoryear{Bonanos et 
al.}{2004}]{2004ApJ...611L..33B} Bonanos A.~Z., et al., 2004, ApJ, 611, L33 
 

\bibitem[]{} Bonnell, I. A., Bate, M. R.,  1994, MNRAS,  271, 999

\bibitem[]{} Bonnell, I. A., Bate, M. R.,  2002, MNRAS,  336, 659

\bibitem[]{} Bonnell, I. A., Bate, M. R.,  Clarke, C. J., \& Pringle, J. E., 2001a, MNRAS, 323, 785

\bibitem[]{} Bonnell I. A., Bate M. R., Vine S. G., 2003, \mnras, 343, 413 

\bibitem[]{} Bonnell, I. A., Bate, M. R., \& Zinnecker, H., 1998, MNRAS, 298, 93

\bibitem[]{} Bonnell I. A., Clarke C. J., Bate M. R., Pringle J. E., 2001, \mnras, 324, 573.

\bibitem[]{} Bonnell I. A., Vine S. G.,  Bate M. R.,  2004, \mnras, 349, 735

\bibitem[]{} Clarke C.J., Bonnell I. A., Hillenbrand L. A., 2000, , in {Protostars and Planets IV} (eds V. Mannings, A. P. Boss and S. Russell), 151.

\bibitem[\protect\citeauthoryear{Clarke \& 
Pringle}{1992}]{1992MNRAS.255..423C} Clarke C.~J., Pringle J.~E., 1992, 
MNRAS, 255, 423 

\bibitem[]{} Davdison K., Humphreys R. M., 1997; ARA\&A, 35, 1

\bibitem[]{} de Wit W.J., Testi L., Palla F., Zinnecker H., 2005, A\&A, in press

\bibitem[]{} Dobbs C.L., Bonnell I.A., Clark P.C., 2005, MNRAS, in press

\bibitem[\protect\citeauthoryear{Edgar \& 
Clarke}{2004}]{2004MNRAS.349..678E} Edgar R., Clarke C., 2004, MNRAS, 349, 
678 

\bibitem[]{} Figer D. F., Najarro F., Morris M., \etal~ 1998, ApJ, 506, 384
 
\bibitem[\protect\citeauthoryear{Figer et al.}{1999}]{1999ApJ...525..750F} 
Figer D.~F., Kim S.~S., Morris M., Serabyn E., Rich R.~M., McLean I.~S., 
1999, ApJ, 525, 750 
 

\bibitem[\protect\citeauthoryear{Garc{\'{\i}}a \& 
Mermilliod}{2001}]{2001A&A...368..122G} Garc{\'{\i}}a B., Mermilliod J.~C., 
2001, A\&A, 368, 122 

\bibitem[\protect\citeauthoryear{Garmany, Conti, \& 
Massey}{1980}]{1980ApJ...242.1063G} Garmany C.~D., Conti P.~S., Massey P., 
1980, ApJ, 242, 1063 

\bibitem[\protect\citeauthoryear{Kroupa, Aarseth, \& 
Hurley}{2001}]{2001MNRAS.321..699K} Kroupa P., Aarseth S., Hurley J., 2001, 
MNRAS, 321, 699 

\bibitem[]{} Lada C. J., Lada, E. 2003, ARA\&A, 41, 57

\bibitem[]{} Larson, R.B., 2003, Rep. Prog. Physics, {\bf 66}, 1651

\bibitem[\protect\citeauthoryear{Mason et al.}{1998}]{1998AJ....115..821M} 
Mason B.~D., Gies D.~R., Hartkopf W.~I., Bagnuolo W.~G., Brummelaar T.~T., 
McAlister H.~A., 1998, AJ, 115, 821 

\bibitem[]{} Massey, P., Penny L. R., Vukovich J., 2002, ApJ, 565, 982

\bibitem[\protect\citeauthoryear{McCaughrean \& 
Stauffer}{1994}]{1994AJ....108.1382M} McCaughrean M.~J., Stauffer J.~R., 
1994, AJ, 108, 1382 

\bibitem[]{} McKee, C.F.,  Tan, J.C., 2003, ApJ, 585, 850

%\bibitem[]{} Mermilliod J.C., 2001, in IAU 200, eds H. Zinnecker and R. Mathieu

\bibitem[]{} Monaghan J. J., 1992, \ARAA, 30, 543.

\bibitem[\protect\citeauthoryear{Preibisch et 
al.}{1999}]{1999NewA....4..531P} Preibisch T., Balega Y., Hofmann K., 
Weigelt G., Zinnecker H., 1999, NewA, 4, 531 

\bibitem[\protect\citeauthoryear{Pringle}{1991}]{1991MNRAS.248..754P} 
Pringle J.~E., 1991, MNRAS, 248, 754 

\bibitem[\protect\citeauthoryear{Rauw et al.}{2004}]{2004A&A...420L...9R} 
Rauw G., et al., 2004, A\&A, 420, L9 

\bibitem[]{} Rauw G., Vreux, J.-M., Gosset E., \etal 1996, A\&A, 306, 771
 
\bibitem[\protect\citeauthoryear{Scally, Clarke, \& 
McCaughrean}{2005}]{2005MNRAS.358..742S} Scally A., Clarke C., McCaughrean 
M.~J., 2005, MNRAS, 358, 742 

\bibitem[\protect\citeauthoryear{Stahler, Palla, \& 
Ho}{2000}]{2000prpl.conf..327S} Stahler S.~W., Palla F., Ho P.~T.~P., 2000, 
prpl.conf, 327 

\bibitem[]{} Wolfire M.G., Cassinelli J.P., 1987, ApJ, 319, 850

\bibitem[]{} Yorke H., Kr\"ugel E., 1977, A\&A, 54, 183

\bibitem[]{} Yorke H., Sonnhalter, C., 2002, ApJ, 569, 846


\end{thebibliography}
\end{document}